\documentclass[11pt]{article}
\usepackage{mymoriond2,epsfig}

\def\be{\begin{equation}}
\def\ee{\end{equation}}
\def\bea{\begin{eqnarray}}
\def\eea{\end{eqnarray}}

\newcommand{\lsim}{
\mathrel{\hbox{\rlap{\hbox{\lower4pt\hbox{$\sim$}}}\hbox{$<$}}}}

\newcommand{\gsim}{
\mathrel{\hbox{\rlap{\hbox{\lower4pt\hbox{$\sim$}}}\hbox{$>$}}}}

\begin{document}
\begin{flushright}
\begin{tabular}{l}
IPPP/07/09\\
DCPT/07/18
\end{tabular}
\end{flushright}

\vspace*{4cm}

\title{\boldmath BASICS OF $D^0$--$\bar D^0$ MIXING}

\author{PATRICIA BALL}

\address{IPPP, Department of Physics, University of Durham,\\
Durham DH1 3LE, England}

\maketitle\abstracts{
Complementing the presentations, at this conference, 
of the first experimental evidence for $D$ mixing found at BaBar and
Belle, I discuss the theoretical status of $D$ mixing.
\\[2cm]
{\it Quasi-Impromptu Talk given at XLIInd Rencontres de Moriond,
  Electroweak Interactions and Unified Theories, La Thuile, Italy,
  March 2007}}

\newpage

The highlight of this year's Moriond conference on electroweak 
interactions and unified theories arguably was the announcement by
BaBar and Belle of experimental evidence for $D^0$-$\bar D^0$ mixing
\cite{staric,flood}, accompanied by an experimental paper \cite{BaBar}
and, less than one week after the event, a theoretical
analysis \cite{fast} -- very likely to be followed by many others. As the
experimental result came as a surprise to everyone, including the
conference organisers, no theoretical talk on the topic had been
organised. I was asked to fill the gap and give a
quasi-impromptu talk on the theory basics of $D$ mixing, whose
written form is presented in these pages. Excellent reviews on the
topic can be found in Refs.~\cite{review1,review2}, and an
enlightening reminder of the importance of charm physics in Ref.~\cite{bigi}.

\begin{figure}[b]
$$\epsfxsize=0.45\textwidth\epsffile{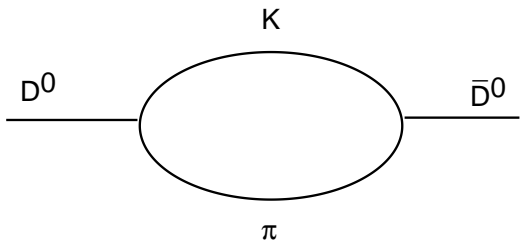}$$
\vspace*{-1cm}
\caption[]{Resonance contribution to $D$ mixing. Figure taken from
  Ref.~\cite{review1}.}\label{fig1}
\end{figure}

In complete analogy to $B$ mixing, $D$ mixing in the SM is due to box
diagrams with internal quarks and $W$ bosons. In contrast to $B$,
though, the internal quarks are down-type. Also in contrast to $B$
mixing, the GIM mechanism is much more effective, as the heaviest down-type
quark, the $b$, comes with a relative enhancement factor
$(m_b^2-m_{s,d}^2)/(m_s^2-m_d^2)$, but also a large CKM-suppression
factor $|V_{ub}V_{cb}^*|^2/|V_{us} V_{cs}^*|^2\sim \lambda^8$, which
renders its contribution to $D$ mixing $\sim 1\%$ and hence negligible.
As a consequence, $D$ mixing is small in the SM, which makes it
very sensitive to the potential intervention of new physics (NP), but on
the other hand, it is also more difficult to accurately calculate the
SM ``background'', as the loop-diagrams are dominated by $s$ and $d$ quarks
and hence sensitive to the intervention of resonances and
non-perturbative QCD, see Fig.~\ref{fig1}. 
The quasi-decoupling of the
3rd quark generation also implies that CP violation in $D$ mixing is
extremely small in the SM, and hence any observation of CP violation
will be a clear-cut signal of new physics, independently of hadronic
uncertainties. 

The theoretical parameters describing $D$ mixing can be defined in
complete analogy to those for $B$ mixing: the time evolution of the
$D^0$ system is described by the Schr\"odinger equation
\begin{equation}
\frac{\partial}{\partial t}\left(\begin{array}{c} D^0(t) \\ \bar
    D^0(t)\end{array}
\right) = - i \left(M - i \frac{\Gamma}{2}\right)
\left(\begin{array}{c} D^0(t) \\ \bar
    D^0(t)\end{array}
\right)
\end{equation}
with Hermitian  matrices $M$ and $\Gamma$. The off-diagonal elements
of these matrices, $M_{12}$ and $\Gamma_{12}$, describe, respectively, 
the dispersive and absorptive parts of $D$ mixing.
The flavour-eigenstates
$D^0=(c\bar u)$, $\bar D^0=(u\bar c)$ differ from the mass-eigenstates
$D_{1,2}$; they are related by
\begin{equation}
|D_{1,2}\rangle = p |D^0\rangle \pm q | \bar D^0\rangle
\end{equation}
with
\begin{equation}
\left| \frac{q}{p}\right|^2 =
\frac{M_{12}^*-\frac{i}{2}\,\Gamma^*_{12}}{ 
M_{12}-\frac{i}{2}\,\Gamma_{12}}\,.
\end{equation}

The basic observables in $D$ mixing are the mass and lifetime
differences of $D_{1,2}$, which are usually normalised to the
average lifetime $\Gamma = (\Gamma_1+\Gamma_2)/2$:
\begin{equation}
x\equiv \frac{\Delta M}{\Gamma} = \frac{M_2-M_1}{\Gamma}\,,\quad
y\equiv \frac{\Delta \Gamma}{2\Gamma} =
\frac{\Gamma_2-\Gamma_1}{2\Gamma}\,.
\end{equation}
While previously only bounds on $x$ and $y$ were known, both BaBar and Belle
have now obtained evidence for a non-vanishing mixing in the $D$
system. BaBar has obtained this evidence from the measurement of the
doubly Cabibbo-suppressed decay $D^0\to K^+\pi^-$ (and its CP
conjugate), yielding
\begin{equation}
y' = (0.97\pm 0.44({\rm stat})\pm 0.31({\rm syst}))
\times 10^{-2},\quad 
x'^2 = (-0.022\pm 0.030({\rm stat})\pm 0.021({\rm syst}))\times
10^{-2},
\end{equation}
while Belle obtains
\begin{equation}
y_{\rm CP} = (1.31\pm 0.32({\rm stat})\pm 0.25({\rm syst}))\times
10^{-2}
\end{equation}
from $D^0\to K^+K^-, \pi^+\pi^-$ and 
\begin{equation}
x = (0.80\pm 0.29({\rm stat})\pm 0.17({\rm syst}))\times 10^{-2},\quad
y = (0.33\pm 0.24({\rm stat})\pm 0.15({\rm syst}))\times 10^{-2}
\end{equation}
from a Dalitz-plot analysis of $D^0\to K_S^0\pi^+\pi^-$.
Here $y_{\rm CP}\to y$ in the limit of no CP violation in $D$ mixing,
while the primed quantities $x',y'$ are related to $x,y$ by a rotation
by a strong phase $\delta$, see below.

CP violation in $D^0\to f $ decays, which is predicted to be extremely small 
in the SM, can be characterised by non-vanishing values of
\begin{equation}
A_M = \left| \frac{q}{p}\right| - 1,\qquad
\phi =  \arg (M_{12}/\Gamma_{12}),
\end{equation}
where $A_M$ measures CP violation in the mixing amplitude, while $\phi$
plays a r\^{o}le in the interference between the decays $D^0\to f$ and
$\bar D^0\to f$. 

\begin{figure}
$$\epsfxsize=0.9\textwidth\epsffile{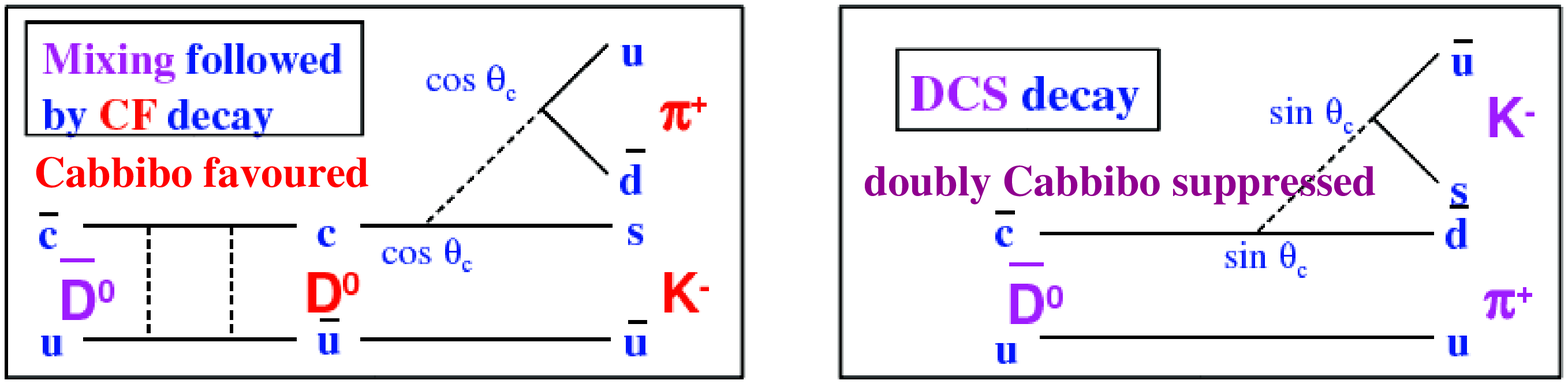}$$
\vspace*{-20pt}
\caption[]{The two amplitudes contributing to the wrong-sign decay $\bar
D^0\to K^-\pi^+$. Figure adapted from Ref.~\cite{flood}.}\label{fig2}
\end{figure}

Various $D^0$ decay channels are sensitive to $D$
mixing. The evidence found by BaBar relies on  $D^0\to
K^+\pi^-$, $\bar D^0\to
K^-\pi^+$, which are ``wrong sign'' decays (the dominant transition is
$c\to s$ and produces a $K^-$ in $D^0$, and a $K^+$ in $\bar
D^0$ decays) and receive contributions from two
amplitudes: a doubly Cabibbo-suppressed amplitude $\bar D^0\to
K^-\pi^+$, i.e.\ $\bar c\to \bar d
\bar u s$, and a two-step process via the oscillation $\bar D^0\to D^0$,
followed by the Cabibbo-favoured process $D^0\to K^-\pi^+$, see 
Fig.~\ref{fig2}.
The relevant point here is that the amplitude with no oscillation is
heavily suppressed which makes it competitive with the oscillated
amplitude. 
The wrong-sign time-dependent decay rate $D^0(t)\to K^+\pi^-$ 
is usually normalised to the
Cabbibo-favoured rate $D^0\to K^-\pi^+$. 
Expanding the ratio of suppressed vs.\ favoured amplitudes 
to second order in $x,y$, one finds
\begin{eqnarray}
\frac{\Gamma(D^0(t)\to K^+\pi^-)}{\Gamma(D^0\to K^-\pi^+)} &=&
\Gamma e^{-\Gamma t} \left[ R_D + \left|\frac{q}{p}\right| \sqrt{R_D}
  (y'\,\cos \phi - x'\,\sin \phi) (\Gamma t) +
  \left|\frac{q}{p}\right|^2 \frac{x'^2+y'^2}{4}\,(\Gamma t)^2\right],
\nonumber\\
\frac{\Gamma(\bar D^0(t)\to K^-\pi^+)}{\Gamma(\bar D^0\to K^+\pi^-)} &=&
\Gamma e^{-\Gamma t} \left[ R_D + \left|\frac{p}{q}\right| \sqrt{R_D}
  (y'\,\cos \phi + x'\,\sin \phi) (\Gamma t) +
  \left|\frac{p}{q}\right|^2 \frac{x'^2+y'^2}{4}\,(\Gamma
  t)^2\right],\nonumber\\[-4pt]\label{9}
\end{eqnarray}
where the overall factor $\Gamma$ ensures the correct normalisation
upon integration over $t$. Here $R_D^{1/2}$ is the modulus of the
ratio of the doubly Cabbibo-suppressed amplitude vs.\ the favoured one
and $x',y'$ contain the effect of the relative strong phase $\delta$ between
the two amplitudes:
\begin{eqnarray}
\frac{A(D^0\to K^+\pi^-)}{A(\bar D^0\to K^+\pi^-)} 
&=& - R_D^{1/2} e^{-i\delta},\label{10}\\
x' & = & x \cos\delta + y\sin\delta,\quad y' = y\cos\delta -
x\sin\delta\,;
\end{eqnarray}
$\delta$ vanishes in the SU(3) limit.
The minus sign in (\ref{10}) originates from the sign of $V_{us}$
relative to $V_{cd}$. Note that the 2nd term in brackets in (\ref{9})
comes from the interference of the two decay amplitudes with and
without mixing. BaBar has obtained $R_D$, $y'$ and $x'^2$ from
the fit of their experimental results to the above formulas in the
case of (a) CP conservation, i.e.\ $|q/p|\to 1$, $\phi\to 0$, and
(b) CP violation, i.e.\ different coefficients $R_{D}^\pm$, $y'_\pm$
etc.\ for $D^0$ and $\bar D^0$ decays. The difference of the latter
proved to be compatible with 0, so there is no evidence for CP violation.

Let us now turn to the theoretical predictions for $x$ and $y$ in the
SM. In terms of hadronic matrix elements, $M_{12}=M_{21}^*$ and
$\Gamma_{12}=\Gamma_{21}^*$
can be expressed as
\begin{eqnarray}
M_{21} & = & \langle \bar D^0 | {\cal H}_{\rm eff}^{\Delta C=2} |
D^0\rangle + P\sum_n \frac{\langle \bar D^0 | {\cal H}_{\rm
    eff}^{\Delta C=1} | n \rangle \langle n | {\cal H}_{\rm
    eff}^{\Delta C=1} | D^0\rangle}{m_D^2 - E_n^2} ,\nonumber\\
\Gamma_{21} & = & P\sum_n \rho_n^{\rm ph.sp} \langle\bar D^0 | {\cal H}_{\rm
    eff}^{\Delta C=1} | n \rangle \langle n | {\cal H}_{\rm
    eff}^{\Delta C=1} | D^0\rangle.
\end{eqnarray}
While the expression for $\Gamma_{12}$ is very similar to that in the
$B$ system, that for $M_{12}$ differs by the contribution of the
second term which is heavily suppressed in $B$ mixing. 
The sum runs over all decay channels of $D^0$; the
contribution to $M_{12}$ includes that of off-shell intermediate
states, while only on-shell states contribute to $\Gamma_{12}$;
$\rho_n^{\rm ph.sp}$ is the corresponding phase-space factor. ${\cal
  H}_{\rm eff}^{\Delta C=2}$ is the local Hamiltonian obtained from
the box diagrams, and includes potential contributions from NP, while
all terms in ${\cal H}_{\rm eff}^{\Delta C=1}$, the Hamiltonian 
describing non-leptonic decays of the $c$ quark, are dominated by SM
contributions (see, however, Ref.~\cite{golowich} for a discussion of
NP effects in  decay amplitudes). Neglecting long-distance
non-perturbative QCD effects, and only including the box diagrams, one
finds \cite{exclusive} $x_{\rm box} = O(10^{-5})$, $y_{\rm box} =
O(10^{-7})$, which is far below the experimental results -- which
indicates that these long-distance effects are extremely important.

There is an extensive literature on
estimating $x$ and $y$ within and beyond the SM, see Ref.~\cite{coll}
for a collection of results. The central problem of all these
calculations is that the $D$ is too heavy to be treated as light and
too light to be treated as heavy. As a consequence, the two approaches
that have been so successful in treating heavy ($B$) and light ($K$)
meson mixing both are not really applicable to $D$ mixing: the
``inclusive'' approach is based on  operator product expansion
and relies on quark-hadron duality. If $\Lambda/m_c$, where $\Lambda$ is a
hadronic scale, is considered a small parameter, $x$ and $y$ can be
expanded in terms of matrix elements of local operators
\cite{inclusive}, and the series can be truncated after a few
terms. Such calculations typically yield
$x,y\,$\raisebox{-3pt}{$\stackrel{<}{\sim}$}$\,10^{-3}$, 
and the result of both BaBar and Belle, $y\sim 10^{-2}$, is
certainly not a generic prediction of such an analysis. In the
``exclusive'' approach~\cite{exclusive,excl-old}, on the other hand, one sums
over intermediate hadronic states, which may be modeled or fit to
experimental data. One crucial observation~\cite{exclusive} is that
$x$ and $y$ are only generated at second order in SU(3) breaking,
which suggests an analysis based on the summation over exclusive
states arranged in SU(3) multiplets. As argued in Ref.~\cite{exclusive},
the main source
of SU(3)-breaking within these multiplets 
is due to phase-space, or rather, the lack
thereof: if the heaviest members of a multiplet are too heavy to be
kinematically
accessible in the decay, they have to be excluded from the sum
over all members of the multiplet (e.g.\ $D\to 4\pi$ is kinematically 
allowed, but $D\to 4K$ is not) and
as a consequence, the cancellation of the sum over all terms, which
yields 0 in the SU(3)-limit, is badly broken. 
The conclusion is that in this way values of $y\sim 10^{-2}$ can be
reached -- which agrees very well with the experimental result and
suggests that these threshold effects may indeed explain
the experimental result. The inclusive approach, on the other hand,
relies on the duality of hadronic and partonic effects, smeared over
sufficiently large energy intervals, and is manifestly insensitive to
threshold phenomena -- and hence likely to be inapplicable to $D$
decays. In the exclusive approach, $x$ can 
be related to $y$ via a dispersion relation;  the
authors of Ref.~\cite{exclusive} find that for $y\sim 1\%$ one 
expects $|x|$ between $0.1\%$ and $1\%$, and  $x$ and $y$ to be of
opposite sign; one should be aware, however, that this calculation is more
model-dependent than that of $y$.

In conclusion, we find that the experimental results on $D$ mixing 
reported by BaBar and Belle at the 2007 Rencontres de Moriond on
electroweak interactions and unified theories present a major step
forward in experimental achievement and analysis. The measured value
of $y\,$\raisebox{-3pt}{$\stackrel{>}{\sim}$}$\,x$
is at the high end of theoretical predictions and indicates
large long-distance contributions, which also impact on $x$, i.e.\ the
short-distance/NP sensitive mass difference. As long as there is no
major breakthrough in theoretical predictions for $D$ mixing, which
are held back by the fact that the $D$ meson is at the same time too heavy and
too light for our current theoretical tools to get a proper grip on
the problem, the long-distance SM contributions to $x$ will completely
obscure any NP contributions and their detection. The observation of CP
violation still presents a theoretically clean way for NP to manifest
itself and it is to be hoped that in the near future, i.e.\ at the $B$
factories or the LHC, at least one of the plentiful opportunities for
NP to show up in CP violation~\cite{opportunities} will be realised.

\section*{Acknowledgments}

I would like to thank M.~Neubert for useful suggestions.


\begin{thebibliography}{99}

\bibitem{staric}
M. Staric, talk given at this conference.

\bibitem{flood}
K. Flood, talk given at this conference.

\bibitem{BaBar}
  B.~Aubert {\it et al.} [BABAR Collaboration],
  arXiv:hep-ex/0703020.

\bibitem{fast}
 M.~Ciuchini {\it et al.}, 
  arXiv:hep-ph/0703204.

\bibitem{review1}
  G.~Burdman and I.~Shipsey,
  Ann.\ Rev.\ Nucl.\ Part.\ Sci.\  {\bf 53}, 431 (2003)
  [arXiv:hep-ph/0310076].

\bibitem{review2}
D.~Asner, review on D mixing in   
W.~M.~Yao {\it et al.}  [Particle Data Group],
  J.\ Phys.\ G {\bf 33} (2006) 1;\\
 K.~Flood,
{\it Prepared for Heavy Quarks and Leptons Workshop 2004, San Juan,
 Puerto Rico, 1-5 Jun 2004}
 I.~Shipsey,
  Int.\ J.\ Mod.\ Phys.\  A {\bf 21} (2006) 5381
  [arXiv:hep-ex/0607070];\\
A.~A.~Petrov,
  Int.\ J.\ Mod.\ Phys.\  A {\bf 21} (2006) 5686
  [arXiv:hep-ph/0611361].

\bibitem{bigi}
  I.~I.~Bigi,
  Int.\ J.\ Mod.\ Phys.\  A {\bf 21} (2006) 5404
  [arXiv:hep-ph/0608073].

\bibitem{golowich}
E.~Golowich, S.~Pakvasa and A.~A.~Petrov,
  arXiv:hep-ph/0610039.

\bibitem{exclusive}
  A.~F.~Falk, Y.~Grossman, Z.~Ligeti and A.~A.~Petrov,
  Phys.\ Rev.\  D {\bf 65}, 054034 (2002)
  [arXiv:hep-ph/0110317];\\
  A.~F.~Falk, Y.~Grossman, Z.~Ligeti, Y.~Nir and A.~A.~Petrov,
  Phys.\ Rev.\  D {\bf 69}, 114021 (2004)
  [arXiv:hep-ph/0402204].

\bibitem{coll}
H.~N.~Nelson,
 in {\it Proc. of the 19th Intl. Symp. on Photon and Lepton
  Interactions 
at High Energy LP99 }, ed. J.A. Jaros and M.E. Peskin,
  arXiv:hep-ex/9908021.

\bibitem{inclusive}
  H.~Georgi,
  Phys.\ Lett.\  B {\bf 297}, 353 (1992)
  [arXiv:hep-ph/9209291];\\
  T.~Ohl, G.~Ricciardi and E.~H.~Simmons,
  Nucl.\ Phys.\  B {\bf 403}, 605 (1993)
  [arXiv:hep-ph/9301212];\\
  I.~I.~Y.~Bigi and N.~G.~Uraltsev,
  Nucl.\ Phys.\  B {\bf 592}, 92 (2001)
  [arXiv:hep-ph/0005089].

\bibitem{excl-old}
J.~F.~Donoghue, E.~Golowich, B.~R.~Holstein and J.~Trampetic,
  Phys.\ Rev.\  D {\bf 33} (1986) 179;\\
F.~Buccella, M.~Lusignoli and A.~Pugliese,
  Phys.\ Lett.\  B {\bf 379} (1996) 249
  [arXiv:hep-ph/9601343];\\
E.~Golowich and A.~A.~Petrov,
  Phys.\ Lett.\  B {\bf 427}, 172 (1998)
  [arXiv:hep-ph/9802291].

\bibitem{opportunities}
  P.~Ball and R.~Zwicky,
  JHEP {\bf 0604} (2006) 046
  [arXiv:hep-ph/0603232];\\
  P.~Ball and R.~Fleischer,
  Eur.\ Phys.\ J.\  C {\bf 48}, 413 (2006)
  [arXiv:hep-ph/0604249];\\
P.~Ball and R.~Zwicky,
  Phys.\ Lett.\  B {\bf 642} (2006) 478
  [arXiv:hep-ph/0609037];\\
  P.~Ball, G.~W.~Jones and R.~Zwicky,
  Phys.\ Rev.\  D {\bf 75} (2007) 054004
  [arXiv:hep-ph/0612081].

\end{thebibliography}
\end{document}